\documentclass[apj]{emulateapj}
\usepackage{mathptmx}
\begin{document}

\title{Gravitational Microlensing and the Structure of Quasar Outflows}
\author{Doron Chelouche\altaffilmark{1,2}}
\altaffiltext{1} {School of Natural Sciences, Institute for Advanced Study, Princeton, Einstein Drive, NJ 08540, USA}
\altaffiltext{2} {Chandra Fellow}
\shortauthors{Chelouche D.}
\shorttitle{Microlensing and the Structure of Quasar Outflows}

\begin{abstract}

We show that invaluable information on the 
structure quasar outflows can be obtained  by considering microlensing 
induced variability of absorption line troughs in lensed
quasars. Depending on the structure and geometry of the outflowing
gas, such extrinsic line variability can be  manifested as changes to the
equivalent width of the line as well as line profile
distortions.  Here we consider several physically  distinct outflow models,  having very similar spectral predictions, and show how ML induced absorption line variability can be used to distinguish between them. Merits of future systematic studies of these effects are exemplified.

\end{abstract}


\section{introduction}

A large fraction of type-I (broad emission line) quasars show gas
outflowing from their centers with velocities in the range
$10^2-10^4~{\rm km~s^{-1}}$ (inferred from the detection of
blueshifted absorption lines; see Crenshaw, Kraemer, \& George 2004 and
references therein). Such flows are believed to be launched from
the central engine of quasars (e.g., Elvis 2000) and may have a
considerable effect on their environment (e.g., King 2003, Scannapieco
\& Ho 2004). 

Physical models for quasar outflows include radiation pressure driven
cloud/wind models (e.g.,  Arav, Li, \& Begelman 1994, Murray, Chiang,
Grossman, \& Voit 1995, Proga, Stone, \& Kallman 2000, and Chelouche
\& Netzer 2001), magnetic winds (e.g., K\"{o}nigl, \& Kartje 1994),
and thermal pressure driven flows (e.g., Woods et al. 1996, Chelouche
\& Netzer 2005). While different models imply very different physical
properties of the outflowing gas (such as location, density, and mass
loss rate), they are all in qualitative agreement with
observations. Thus, despite some 30 years of research little is known
with confidence about the physics of such flows.  

In Chelouche (2003) we have shown that the effect of strong
gravitational lensing can be used as a probe for the structure of
quasar outflows over angular scales comparable to the separation between
quasar images. The proposed method was useful in cases where
the effects of gravitational microlensing (ML) may be neglected. The focus
of this work is to consider cases where ML effects are likely to
dominate (e.g., Ofek \& Maoz 2003) and investigate how they can be used to shed
light on the physics of quasar outflows. Specifically, we focus on the
information that can be gained by studying the effects of
ML on absorption line profiles in
lensed quasars. Such effects are potentially  significant
(e.g., Hutsemekers 1993, Lewis \& Belle 1998) yet the implications for
studying quasar outflows in the framework of realistic physical models
have not been addressed. 

This paper is organized as follows: In \S 2 we discuss 
current paradigms in flow modeling and elaborate on the formation of
absorption line profiles in quasars. Section 3 describes the effect of
ML on the shape of line troughs for several flow models. Conclusions
follow in  \S 4. 

\section{The formation of line troughs}

\begin{figure*}
\centerline{\includegraphics[width=17.0cm]{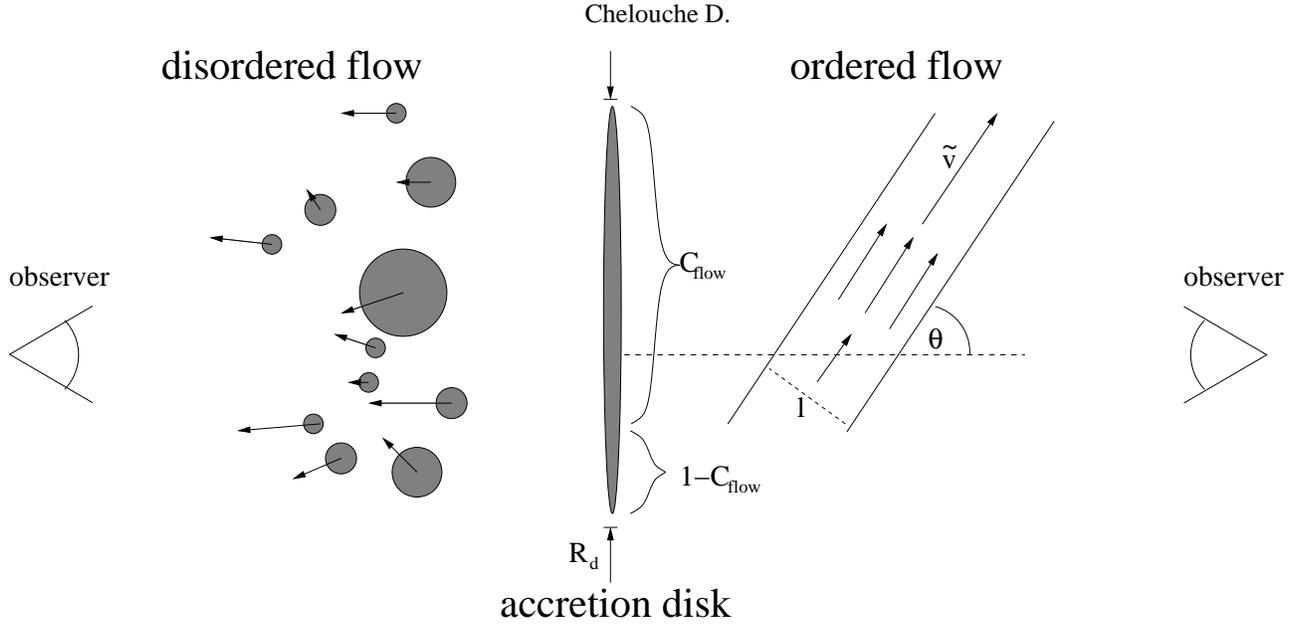}}
\caption{Ordered (A99) vs. 
  disordered (de Kool et al. 2002) flow geometries. Ordered flows (right panel) are
  envisioned as monotonically accelerating gas streams (with thickness
  $l$) that cross our line-of-sight  with inclination angle
  $\theta$. Disordered flows (left panel) may consist of numerous clouds randomly
  distributed in real space and in velocity space, each having its
  properties (e.g., optical depth) drawn from the same parent
  population. We study cases where the total line-of-sight covering
  factor of the flow, $C_{\rm flow}$, is equal or smaller than unity (see \S 3.1).} 
\label{arav}
\end{figure*}

The formation of absorption line profiles depends on the
projected 2D properties of the absorbing gas (such as velocity,
temperature, and ionization structure) and on the continuum
source (e.g., emissivity and size). The residual flux under the line
formed by a flow which is located between us and the continuum emitting source satisfies,  
\begin{equation}
F_{\rm obs}(v)=\int d{{\bf  R}^2} \mu({\bf R}) I({\bf R}) e^{-\tau({\bf R},v)},
\label{form}
\end{equation}
where ${\bf R}$ is a vector marking  the projected
position over the plane of the sky,  $I$ the emitted flux per unit projected
area and is a slowly varying function of wavelength in the vicinity of
the line (for the sake of simplicity we neglect the possible
contribution from emission lines), $\mu({\bf R})$ the
(de-)magnification pattern due to ML 
by stars in the lensing galaxy (see below), and $\tau({\bf R},v)$ the
position and velocity dependent optical depth which depends on the properties of the outflowing gas. 

An equivalent, observationally inclined expression for the residual flux is 
\begin{equation}
F_{\rm obs}(v)=F \left \{ \left [1-C(v) \right ]+C(v)e^{-\tau(v)}
\right \},
\label{lf}
\end{equation}
where the unabsorbed flux emitted by the source,    
\begin{equation}
F=\int d{\bf R}^2 I({\bf R}).
\label{ftot}
\end{equation}
$\tau(v)$ and $C(v)$ are the {\it observed} velocity dependent
optical depth and covering factor, respectively (e.g., Arav, Korista,
\& de Kool 2002). A major breakthrough in flow research was the
realization that the shape of the rest ultraviolet (UV) line troughs
(e.g., \ion{C}{4}\,$1549{\rm \AA}$) is determined by velocity
dependent partial covering effects (e.g.,  
Arav et al. 1999; hereafter A99). Both $\tau(v)$ and $C(v)$ can, in
principal, be deduced from observations and used to infer on the
physics of the flows (i.e., on $\tau({\bf R},v)$, A99). In practice, this
inversion problem is under-constrained and physically  distinct
outflow models give similar line profiles. This model degeneracy is
explained below. 

\begin{figure}
\plotone{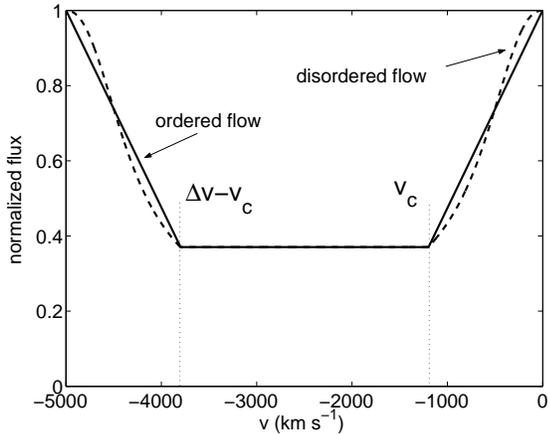}
\caption{The absorption line profiles predicted by
  ordered and disordered flows. Note the similarity between the lie
  troughs predicted by the different models. Here we assume:
  $C_{\rm flow}=1$, $v_c=1200~{\rm km~s^{-1}}$, $C_{\rm max}=0.63$, $\Delta
  v=5000~{\rm km~s^{-1}}$, $\tau \gg 1$ for the ordered flows and
  $p=-0.62$ and  $\tau_{\rm max}(v<1200~{\rm km~s^{-1}})\propto
  v^{2.3},~\tau_{\rm max}(v \geq 1200~{\rm km~s^{-1}} ~\&~ v<3800~{\rm
    km~s^{-1}})= 10~\tau_{\rm max}(v\geq 3800~{\rm km~s^{-1}})\propto
  v^{-2.3}$ for the disordered flows.} 
\label{arav2}
\end{figure}

\subsection{Ordered flows}

This model, as proposed by A99, assumes that the flow is inclined to our
line of sight by angle $\theta$ and is monotonically accelerated across the projected surface of the continuum source (see the right panel of figure 1). This means that every
projected velocity interval, $\delta v$ ($=\delta \tilde{v} {\rm cos}
\theta$, where $\tilde{v}$ is measured along flow lines; see figure
1), corresponds to a different region in the source plane. (Note,
however, that in this model there is a {\it range} of velocities at
every location ${\bf R}$). Here we consider a simplified version of
the model where the velocity gradient, $d\tilde{v}/d\tilde{r}$ ($\tilde{r}$ is
measured along flow lines) and the flow density are
constant.  For a flow whose line troughs extend from $v=0~{\rm km~s^{-1}}$
(corresponding to the rest wavelength of the line) to some maximum
velocity, $\Delta v$, we obtain   
\begin{equation}
\begin{array}{l}
\tau(v)={\rm const.} \\
C(v)= C_{\rm flow}\times
\left \{ \begin{array}{ll}
\displaystyle C_{\rm max} v/v_c; & v<v_c \\
\displaystyle C_{\rm max}; & v>v_c~{\rm and}~\Delta v-v>v_c\\
\displaystyle C_{\rm max} (\Delta v-v)/v_c; & \Delta v-v<v_c \\
\end{array}
\right .
\end{array}
\label{flow}
\end{equation}
where $C_{\rm max}=l{\rm cos}\theta/R_d$
is the maximum covering factor, and $v_c=(d\tilde{v}/d\tilde{r})(l/{\rm tg}\theta)
{\rm cos}\theta$. $C_{\rm flow}$ is the fraction of the ionizing source which is covered by the outflow. An example of the absorption line profile produced
by such a flow is shown in figure 2.

\subsection{Disordered flows}

A different model, suggested by de Kool et al. (2002; cf. Lewis \&
Belle 1998), considers a flow comprised of numerous small (compared to
$R_d$) clouds that are {\it randomly} distributed in real and in
velocity space. Each cloud from an ensemble of clouds traveling with
velocity $v$ (there are many such ensembles covering the full velocity
range, $\Delta v$) has a different optical depth drawn from a powerlaw
distribution with index $p$  and a maximum
optical depth distribution, $\tau_{\rm  max}(v)$ (see equation
9 in de Kool et al. 2002). The residual flux (see their equation 13) is 
\begin{equation}
F_{\rm obs}(v)=F \frac{(p+1)\Gamma[p+1]}{\tau_{\rm max}(v)^{p+1}}P
\left [ p+1, \tau_{\rm max} (v) \right ].
\label{dek}
\end{equation}
$\tau_{\rm max} (v),~p$ are not known a-priori, but can be deduced 
from observations, and $\Gamma,~P$ are the complete and incomplete gamma
functions, respectively (see de Kool et al. 2002 for a detailed layout
of the formalism). Figure 2 demonstrates
that, the absorption line profiles from ordered and disordered flows
can be very similar for a suitable choice of parameters, and cannot therefore be easily distinguished.  Below 
we wish to examine whether this degeneracy can be 
removed by considering the effects ML has on such line profiles in lensed quasars (e.g., Hutsemekers 1993). 

\section{Distortion of line troughs by Microlensing}

\begin{figure}
\plotone{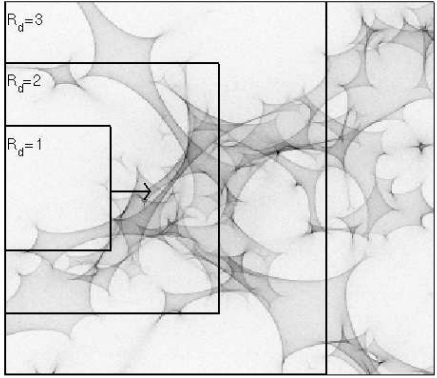}
\caption{The magnification map, $\mu({\bf R})$, used here to
  calculate the distortion of the absorption line profiles formed by quasar flows (equation
  1). Dark shades correspond to higher magnifications. A web of
  caustics is clearly shown resulting from the combined effect of many
  stars in the lensing galaxy. Here we assume a microlensing
  mass surface density, $\kappa=0.55$ with no external shear ($\gamma=0$). 
  The rectangles mark the initial position and size, $R_d$, of the
  quasar continuum emitting region. $R_d=1$ corresponds to roughly three
times the Einstein radius of one solar mass star ($\simeq 3\times10^{16}~{\rm cm}$ for a typical lensed system where the quasar, lens redshifts are $2,~0.5$, respectively; cf. Treyer \&  Wambsganss 2004).} 
\label{ml}
\end{figure}

Recent studies have shown that ML can induce significant ($>0.1$
magnitude) fluctuations in the light-curves of quasars (e.g.,
Irwin et al. 1989, Wo\'{z}niak et al. 2000, Ofek \& Maoz 2003). This
effect results from 
the Einstein radius of stars in the lensing galaxy being comparable to
the size of the quasar continuum source (e.g.,  Schneider, Ehlers, \&
Falco 1992 and references therein). It has been suggested that ML
can be used to explore the central engines (e.g.,
Agol \& Krolik 2000) and the broad emission line regions in quasars (e.g., Abajas et al. 2002, Chartas et al. 2004, Lewis \& Ibata 2004, Richards et al. 2004). In  particular, Hutsemekers (1993) and Lewis \& Belle (1998) have shown
that single-caustic crossing events can distort the shape of
absorption line profiles.  Such distortions result when the
magnification pattern is non-uniform across the continuum source
(i.e.,  $\mu({\bf  R})\ne$const in equation \ref{form}) and can be
manifested as equivalent width (EW) changes (cf. Lewis \& Ibata
2003), absorption line profile distortions, and their combination. 

Qualitatively, pure EW changes occur when a continuum region which is
occulted by the absorbing gas (e.g., a cloud) is (de-)magnified with
respect to the non-occulted part of the source. Shape distortions
occur when different parts of the flow traveling with distinct
velocities cover separate regions of the continuum source which
are subjected to different ML magnifications. A more quantitative treatment
requires the calculation of realistic  magnification patterns and flow
models (e.g., Kochanek 2004). The magnification map (source plane), $\mu({\bf R})$ (see figure 3) was calculated using a ray-shooting algorithm similar to that which is described in
Wambsganss (1999). Here consider a canonical case where the redshift of the lensed quasar, $z_s=2$, and the lensing galaxy is at, $z_l=0.5$ (such values are typical of strongly lensed systems; see the CfA-Arizona Space Telescope Lens Survey\footnote{See http://cfa-harvard.edu/castles}; M\~{u}noz et al. 1998). We assume that the dimensionless surface mass density, $\kappa=0.55$ (see Schneider, Ehlers, \& Falco 1992), comprised of equal mass stars whose Einstein radius is $\sim 3\times10^{16}~{\rm cm}$ for the chosen parameters (see figure 3). We assume no external shear. 

As we shall see below, it is instructive to consider three length scales of the problem: The size of the emission region, $R_d$, the auto-correlation length scale of the magnification pattern, $R_\mu$, and the auto-correlation length scale of the flow velocity field, $R_v$. The latter scales are derived from the autocorrelation function, 
\begin{equation}
A_X(r)\propto \int dr' X(r-r') X(r'),
\label{amu}
\end{equation}
where $X$ is the property in question, and $R_X$ is defined as the point beyond
which $A_X$ drops by an $e$-fold. Here $r$ is measured in the mean (over the surface of the emission region) direction of the velocity gradient of the flow. Below we consider simplified models pertaining to the simplified versions of ordered and disordered flows and calculate the effect of ML on the absorption line profiles. We consider cases where ML effects are likely to be significant (e.g., producing detectable variations in the light-curve of the quasar), i.e., where $R_\mu \sim R_d$.

\subsection{Ordered flows}

Here we consider cases which correspond to ordered flows ($R_v\simeq R_d$). We first assume that the flow completely covers the continuum source ($C_{\rm flow}=1$). The effect of ML on the absorption line profiles formed by such flows is shown in figure 4. Our primary conclusion is that line profile distortions can be considerable ($>10\%$ flux changes; see also \S 3.4) and show a  clear distortion pattern extending over the full velocity range covered by the flow. This pattern reflects the correlation between the flow velocity and the magnification pattern (both of which are correlated with ${\bf  R}$).  This result is very different from that of Lewis \& Belle (1998) and from that which corresponds to disordered flows (see below) where line profile distortions occur over small velocity intervals. It is also noted that, for this flow configuration, no significant ML induced EW changes take place. This property is common to all lines that are formed by flows for which $\tau({\bf R},v)=\tau(v)$, i.e., whose properties (such as temperature or ionization) do not depend significantly on ${\bf R}$ (see appendix).

The magnitude and shape of the distortions depend on the flow properties, $\tau(v),~C(v)$. Specifically, larger line optical depths correspond to larger distortion amplitudes ($10\%-30\%$ for $\tau=0.5-3$ respectively, see figure 4a). This is also the case for larger covering fractions (shown in figure 4a as a function of $C_{\rm max}$). This results from the fact that the lower residual flux in such cases  is emitted by an overall smaller region of the continuum source. Such regions are likely to experience larger amplitude (de-)amplifications. Conversly, larger continuum emitting regions (or denser caustic maps due to
a higher stellar density in the lensing galaxy; e.g., Wambsganss 1992)
result in lower amplitude, more complex line distortions (see figure
3b where different source sizes where used in each case). 

In terms of $R_d, R_\mu,$ and $R_v$ this behavior can be understood as
follows: When $R_\mu$ is of the order of $R_d$ ($\sim R_v$, for ordered flows), large (de-)magnifications and considerable line  distortions and EW variations are expected. For the specific example considered here, $R_\mu\simeq 0.5$ (in units of $R_d$)\footnote{Care must be taken when calculating $R_\mu$ since the clumping of caustics in the  magnification pattern  is a highly non-linear effect. Here $R_\mu$ is averaged over the entire magnification map}. (Note that $R_\mu$ is sensitive to the value of $\kappa$ and may vary by a large factor for $\kappa$ in the range $0.2-0.8$; see e.g., figure 1 in Treyer \& Wambsganss 2004). For $R_\mu \ll R_d$ (see for example figure 4b where a source of size $R_d=3$ was assumed), different parts of the continuum emitting region are (de-)magnified by
uncorrelated magnification patterns  so that line variations would
have a more complex, multi-humped pattern and low resolution spectroscopy is likely to find an overall suppression of the effect. On the other extreme where $R_\mu \gg R_d$, line distortions would also be suppressed as different parts of the continuum source are magnified by more or less the same factor (though long-term ML induced light-curve fluctuations may still be considerable; see also \S3.4).

Until now we have considered cases where $C_{\rm  flow}=1$. Nevertheless, flows may not fully cover the entire
emission region (e.g., Chelouche  \& Netzer 2005 and references
therein. In such cases ($C_{\rm flow}<1$ and, effectively, $\tau({\bf R})\ne$const.) considerable EW changes may occur, as demonstrated in the top-left panel of figure 5 for the case where $C_{\rm flow}=0.7$. These changes would be statistically maximized if the non-occulted section of the continuum emitting region is continuous and $R_\mu \simeq R_d$. We further discuss this issue in \S 3.3. 

\begin{figure*}
\centerline{\includegraphics[width=20.0cm]{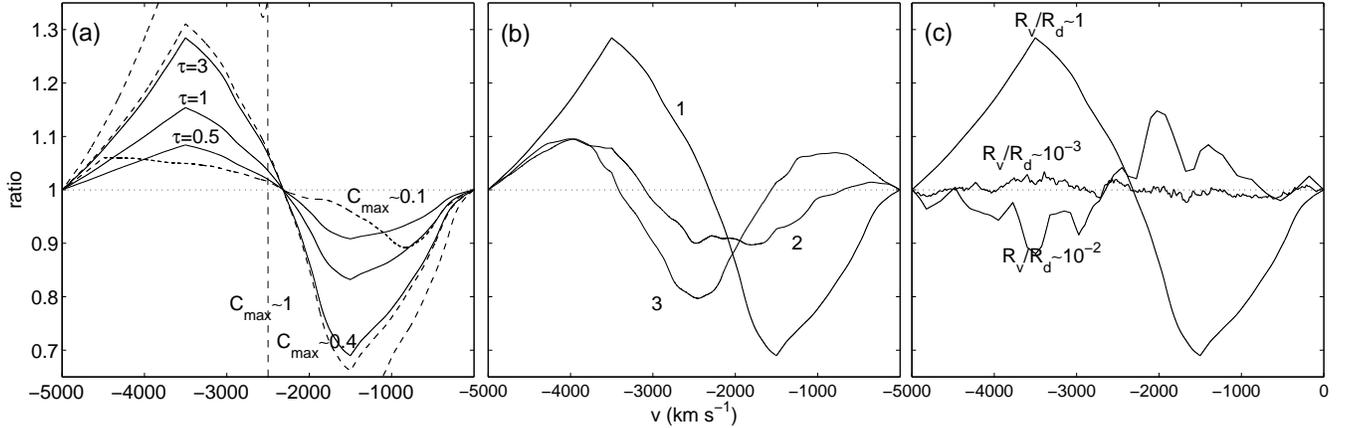}}
\caption{ML induced line distortions shown as the ratio of the
  distorted profile to the un-distorted one ($C_{\rm flow}=1$ is assumed). (a)  More optically thick
  lines (solid lines) or those with a larger covering factor (dashed
  lines) are more distorted by ML effects since the the residual flux is emitted by
  more localized regions of the source. (b) Larger sources ($R_d \gg R_\mu$) will be
  microlensed  by a more complex and less correlated  caustic  pattern
  and the resulting line distortions are less prominent [sources of
  size $R_d=1,2,3$ (in fiducial units) correspond to magnitude fluctuations, $\Delta
  m=0.3, 0.1, 0.03$, respectively]. (c) Line profiles of highly
  disordered flows ($R_v/R_d\ll 1$; see text) are less 
  distorted since the magnification and velocity are poorly
  correlated.} 
\label{dist}
\end{figure*}

\subsection{Disordered flows}

Here we consider the distortion of absorption line profiles formed by disordered flows for which $R_v \ll R_d$. We first consider cases where $C_{\rm flow}=1$ and model those in the following way:  An ordered flow is first constructed using the A99 formalism. The flow is then divided into sections covering each a velocity interval $\delta v$ across (there are $\Delta v/\delta v\simeq R_d/R_v$ such sections). A disordered flow is then obtained by randomly shifting the positions of its sections. The shape of the line profile is conserved  by this
shuffling procedure. Thus, $R_v$ is essentially the size of individual sections so that $R_v/R_d\simeq \delta v/\Delta v$. For example, a model with $R_v/R_d\simeq \delta v/\Delta v=0.3$ can be envisioned as a three stream flow similar to the one conceived by A99 (see their figure 10).

Figure 4c demonstrates that the line profiles formed by less ordered
flows have smaller distortion amplitudes and that the distortion pattern is
more complex, having many low amplitude humps. This results
from the fact that the velocity is poorly correlated with ${\bf R}$ and, hence, with the magnification pattern, For $R_v \ll R_\mu$ line distortions would be negligible even though  the total  ML (de-)magnification of the continuum source is large and continuum fluctuations substantial.  In cases where $R_v \sim R_\mu \ll R_d$, the
effect of gravitational microlensing on the quasar lightcurve as well
as on the line profiles would be negligible. 

We note that although considerable line distortions are unlikely to
occur for highly disordered flows, it is still possible for such lines to
exhibit EW variations if a considerable and continuous region of the
emitting source remains non-occulted by the flow ($C_{\rm
  flow}<1$). EW changes  would then be qualitatively similar to those shown in figure 5 although  no ML induced line profile distortions would be observed in such cases.

\subsection{Time variability}

\begin{figure*}
\centerline{\includegraphics[width=18.0cm]{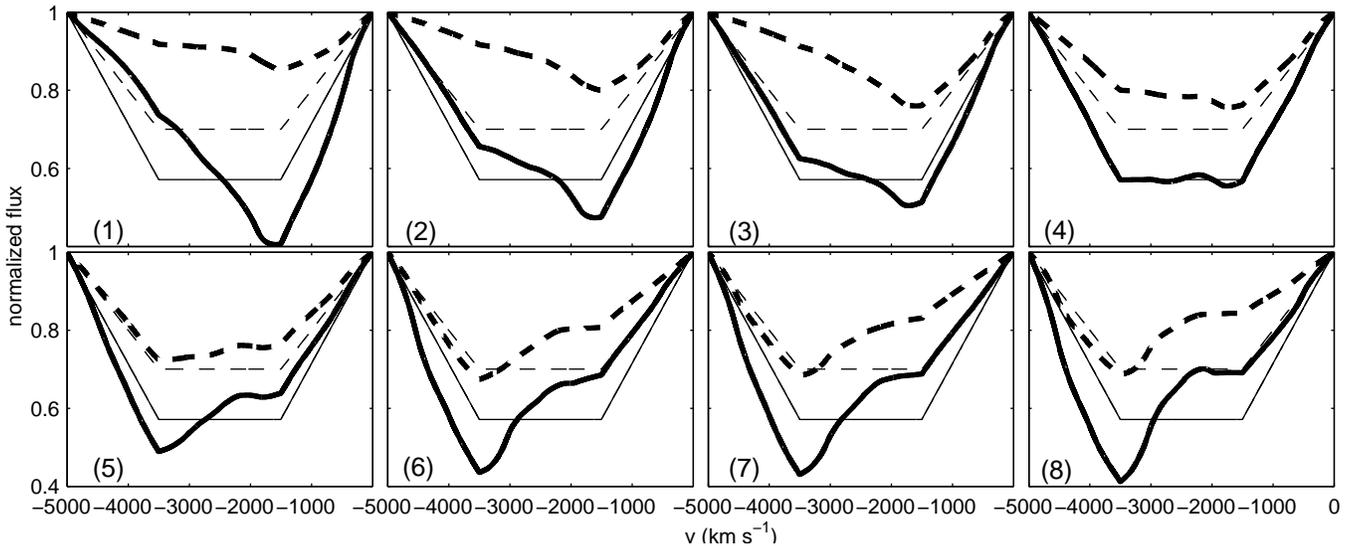}}
\caption{Evolution of the line profiles for an ordered flow
  (assuming an $R_d=1$ continuum source). Solid (dashed) lines assume
  full (70\%) coverage of the continuum source (see
  text). Considerable line 
  distortions are evident in both cases. Significant EW variations are
  expected for a flow that continuously covers only part of the
  continuum source (compare panels 1 and 5). In this example, a hump
  in the line profile seems to propagate with time from high to low
  velocities as a web of caustics (shown at the center of figure 3)
  sweeps across the source.}
\label{var}
\end{figure*}

ML induced effects in the lightcurves of quasars may vary considerably with time (e.g., Schneider, Ehlers, \& Falco  1992, Ofek \& Maoz 2003). This results from the motions of the microlensing stars within the potential well of the lensing galaxy, and the relative peculiar velocity between the quasar, the lens, and the observer.   The relevant timescale for such variations is, roughly, the caustic-crossing timescale which depends on $R_d$ and the peculiar velocity of the galaxy/stars. Typical timescales are of order a few
months to years (e.g., Schneider, Ehlers, \& Falco 1992) and are therefore longer then typical dynamical timescales (e.g., Chelouche 2003). Here we wish to examine how time-dependent ML effects reflect on the observed absorption line profiles. We model such time evolution by artificially shifting the continuum emitting region (with $R_d=1$) across the magnification map (from left to right; see figure 3) and assume a stationary flow. 

Examples for the time-dependent variations of the line profiles from ordered flows are shown for full ($C_{\rm flow}=1$) and partial coverage ($C_{\rm flow}=0.7$) models.  Figure 5 demonstrates that line distortions evolve with time and a  pattern is revealed whereby the dip at low velocities gradually shifts to higher velocities as the central caustic pattern (figure 3) sweeps across the source.  As expected, the EW of the lines remains unaffected by line distortions for a full coverage of the source yet considerable EW changes do occur for the partial covering model (compare panels 1 and 5 in figure 4).  

Significant EW variations of the line troughs can also occur for the case of highly disordered flows provided a continuous section of the emission region remains non-occulted. We note however that in such a case, significant line distortions would be absent.

\subsection{Line profile variation probability}

\begin{figure}
\plotone{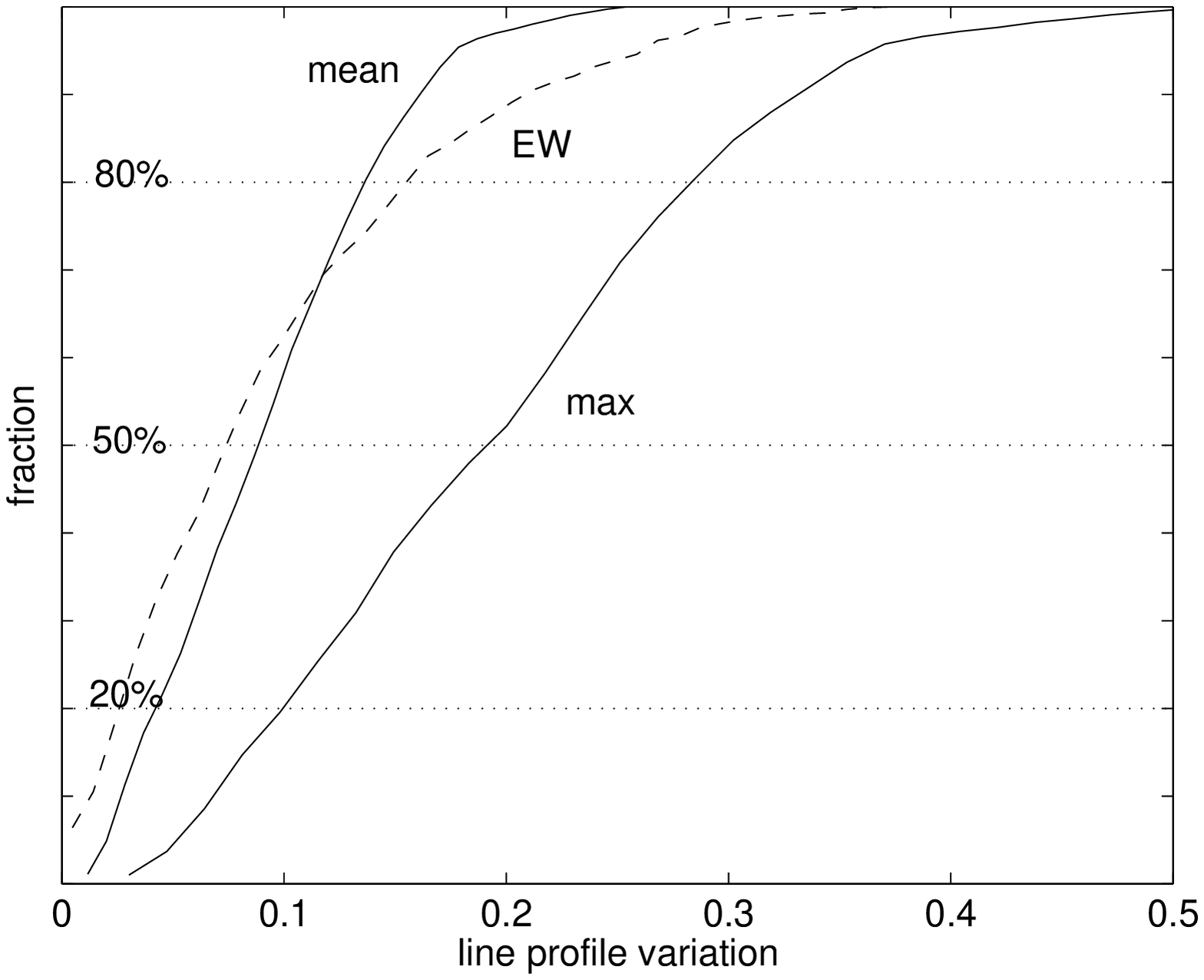}
\caption{The variation probability of the absorption line profiles for a given flow model and magnification map (a source of size one $R_d$ is assumed). Three variation measures are defined: The mean variation is defined as  $\left < \vert \tilde{\phi (\nu)} \left [ \phi(\nu)^{-1}-1 \right ] \vert \right >_\nu $ where $\phi(\nu),~\tilde{\phi(\nu)}$ are the non-distorted, distorted line profiles, respectively. The maximum variation is defined as  ${\rm max} \left \{ \vert \tilde{\phi (\nu)} \left [ \phi(\nu)^{-1}-1 \right ] \vert \right \}$. The EW variation is simply the relative EW difference between the ML and the non-ML  troughs. Solid lines correspond to a model with $C_{\rm flow}=1$ (we note that no significant EW variations occur in this case). Large 10\% mean distortions (solid line) are common while maximum  distortions to the line profile exceed 20\% in most cases. The probability to observe a given EW variation was calculated for a similar flow but with $C_{\rm flow}=0.7$ (shown in dashed line).  Variations of the EW by more than 10\% are frequent.}
\label{dist_p}
\end{figure}

ML induced distortions of absorption line profiles are of a statistical nature related to the position of the quasar over the magnification map. An interesting quantity is therefore the probability to observe some variation at any given time. To calculate this, we have randomly positioned the quasar over the magnification map (assuming a $R_d=1$ emission region) and calculated the line variation for each case. We define the following quantities: The mean distortion amplitude over the line profile, the (local) maximum distortion amplitude, and the relative EW change (see the caption of figure \ref{dist_p} for the proper definitions). 

We first consider ordered flows where we find considerable variations of the line profiles most of the time. For the case considered here,  the mean distortion amplitude exceeds $\sim 10$\% most of the time (with mean distortions exceeding $\sim 15$\% some 20\% of the time). Maximum distortions of the line profiles over localized velocity intervals exceed 20\% most of the time and may be as large as 40\% in some cases. Qualitatively similar results are  obtained when assuming partial line-of-sight coverage although here considerable EW changes may occur (for the example shown here $>10\%$ variations are frequent). 

Very low distortion amplitudes are expected for highly disordered flows. Nevertheless, EW variations will occur if a continuous region region of the continuum source is not occulted by the flow (the result would be similar to that shown in figure 5 for the same covering factor).

Our calculations show that there is no significant correlation between the continuum magnification (i.e., its light-curve) and the variation amplitude of the line profile. This is not unexpected given the fact that line variations require the magnification pattern to vary across the emission region while the total magnification does not (in fact, a coherent magnification or de-magnification pattern would produce the largest continuum variations). 

\section{Conclusions \& Summary}

We have shown that the effect of ML on the absorption line profiles can be utilized to investigate the structure and dynamics of quasar outflows and remove the degeneracy between different outflow models when only one dimensional spectra is considered. Nevertheless, this requires the identification of ML induced line distortion events which may be complicated by the fact that the line profiles of different quasar images may differ due to reasons related to the flow (e.g., Chelouche 2003, Green \& Aldcroft 2003, Ellison et 
al. 2004) or due to incorrect subtraction of light contribution from other
sources (e.g., the lens). Spatially resolved spectral monitoring of lensed quasars is required to elevate some of the difficulties in identifying ML induced line variations.

In cases where line variations are predominantly due to ML then  the implications of the various distortion patterns for the structure and geometry of quasar outflows can be summarized as follows:
\begin{itemize}
\item
Significant line EW variations are observed: A localized, continuous
region of the continuum source is not covered by the outflowing gas. 
\item
Significant line profile distortions are observed and those are well correlated with velocity:  The flow covers most of the continuum source and has a relatively ordered velocity structure. 
\item
Low amplitude, complex distortion pattern is observed: The flow is either
disordered or the magnification pattern highly complex (in 
the latter case ML induced light-curve fluctuations would be suppressed too; e.g., Wambsganss 1992).
\end{itemize}

Most promising candidates for observing  line profile distortions are
lensed quasars showing considerable extrinsic flux variations (i.e., $R_d \simeq R_\mu$) and having non-black, saturated line troughs.  Of particular interest are those quasars having low velocity flows in which case
$\tau(v)$ and $C(v)$ can be independently deduced  (e.g., Arav,
Korista, \& de Kool 2002) and used to put further constraints on 
models. Spatially resolved spectroscopy of two or more quasar images is required to distinguish between extrinsic and intrinsic variations of the absorption troughs (in the latter case, changes would appear in all images after correcting for time-lag effects).

The results from a systematic study designed to look for
absorption line variations in lensed quasars may have far-reaching
implications for understanding the micro-physics of the outflowing gas. This would allow us to derive better constraints on the mass loss rate from quasars and, in turn, to understand their interaction with the environment.

\acknowledgments
The author thanks E. O. Ofek, O. Shemmer, and S. Kaspi for many
fruitful discussions. An anonymous referee is thanked for 
valuable comments. This work is supported by NASA through a Chandra Postdoctoral Fellowship award PF4-50033.

\appendix

Assume that the location and velocity dependent optical depth may be
written in the form
\begin{equation}
e^{-\tau({R},v)}=\eta( {\bf R}) \zeta(v).
\label{etau}
\end{equation}
In this case,
\begin{equation}
\displaystyle {\rm EW}=\frac{\int dv \int d{\bf R}^2 \mu({\bf R})
  I({\bf R}) \left [1-
   \eta( {\bf R}) \zeta(v) \right ]}{\int
  d{\bf R}^2 \mu({\bf R}) I({\bf R})}=\int dv \left [ 1
  -\frac{\zeta(v) \int d{\bf R}^2 \mu({\bf R})
  I({\bf R})\eta({\bf R})}{\int 
  d{\bf R}^2 \mu({\bf R}) I({\bf R})} \right ].
\label{ew1}
\end{equation}
Clearly, if $\eta({\bf R})=$const. then the EW does not depend on the
magnification $\mu$ and is therefore conserved. In case $\eta$ is a
non-trivial function of $R$ then the EW may change as $\mu$ evolves. The  magnitude of these changes depends on the specific form of $\tau({\bf R})$. For instance, when $\tau({\bf R})$ is a highly and rapidly variable function of ${\bf R}$ then effectively non-occulted regions in the disk (where $\tau({\bf R})\simeq 0$) are distributed throughout the disk and are likely to experience different magnifications so that the mean effect is suppressed.

\end{document}